\documentclass[prl, twocolumn]{revtex4-1}
\usepackage{amsmath}
\usepackage{latexsym}
\usepackage{amssymb}
\usepackage{mathrsfs}
\usepackage{bm}
\usepackage{indentfirst}
\usepackage{natbib}
\usepackage{fancyhdr}
\usepackage{graphicx}
\usepackage{relsize}
\usepackage[a4paper, margin=1.6cm,top=3.5cm, bottom=2.5cm, includefoot]{geometry}
\usepackage{epstopdf}
\usepackage{comment}
\usepackage{wrapfig}
\usepackage{placeins}
\usepackage{float}
\usepackage{sidecap}
\usepackage{lmodern}
\usepackage[T1]{fontenc}
\usepackage{verbatim}
\usepackage{enumerate}
\usepackage{setspace}
\usepackage{color}
\usepackage{url}

\def\onedot{$\mathsurround0pt\ldotp$}
\def\cdddot#1{
  \mathbin{\vcenter{\baselineskip.67ex
    \hbox{\onedot}\hbox{\onedot}\hbox{\onedot}%
  }}%
}
\def\fdot#1{
  \mathbin{\vcenter{\baselineskip.6ex
    \hbox{\onedot}\hbox{\onedot}\hbox{\onedot}\hbox{\onedot}%
  }}%
}
\usepackage{appendix}
\usepackage{mathtools}

\begin{document}

 \title{Inverse problem for multi-body interaction of nonlinear  waves}

\author{Alessia Marruzzo$^{1}$, Payal Tyagi$^{1}$, Fabrizio Antenucci$^{1}$, Andrea Pagnani$^{2,3}$,  Luca Leuzzi$^{1,4}$}

\affiliation{$^1$ Soft and Living
  Matter Lab., Rome Unit of CNR-NANOTEC, Institute of Nanotechnology, National Research Council of Italy,
  Piazzale Aldo Moro 5, I-00185, Rome, Italy \\
  $^2$ Department of Applied Science and Technology, Politecnico di Torino, 10129, Torino, Italy \\ $^3$ 
  Human Genetics Foundation, Molecular Biotechnology Center, 10126 Torino, Italy \\$^4$ Dipartimento di Fisica, Universit\`a {\em Sapienza},
  Piazzale Aldo Moro 5, I-00185, Rome, Italy  }

\begin{abstract}
The inverse problem is studied in multi-body systems with nonlinear
dynamics representing, e.g., phase-locked wave systems, standard
multimode and random lasers.  Using a general model for four-body
interacting complex-valued variables we test two methods based on
pseudolikelihood, respectively with regularization and with
decimation, to determine the coupling constants from sets of measured
configurations. We test statistical inference predictions for
increasing number of sampled configurations and for an externally
tunable {\em temperature}-like parameter mimicing real data noise and
helping minimization procedures. Analyzed models with phasors and
rotors are generalizations of problems of real-valued spherical
problems (e.g., density fluctuations), discrete spins (Ising and
vectorial Potts) or finite number of states (standard Potts):
inference methods presented here can, then, be straightforward applied
to a large class of inverse problems. The high versatility of the
exposed techniques also concerns the number of expected interactions:
results are presented for different graph topologies, ranging from
sparse to dense graphs.
\end{abstract}

\maketitle 
Multi-body inference turns out to be essential whenever
non-linear response is crucial for a system properties. Light mode
interaction in ultra-fast multimode lasers
\cite{SargentIII78,Haus00,Gordon02,Katz06,Antenucci15c,Antenucci15d},
random lasers \cite{Lawandy94,Cao98,Wiersma08,Antenucci15a},
multi-variable clause constrained problems \cite{Monasson97,Mezard02},
error correcting codes \cite{MacKay03,Mezard09}, effective interaction
among density fluctuations in heterogeneous frustrated glassy systems
\cite{Kirkpatrick89,Crisanti92,Goetze09,Franz11,Caltagirone12,Ferrari12}
and fish shoals behavior \cite{Katz11,Herbert11} are significant
diverse examples of direct problems in which nonlinearity plays a
non-perturbative role in determining the system behavior.
Nevertheless, to our knowledge, not many studies of the inverse
problem have been performed so far in the field. In this work we aim
at filling in this gap presenting a detailed analysis based on
pseudolikelihood maximization (PLM) techniques for the statistical
inference in models with multi-body interactions.

Inverse problems consist in determining the interaction couplings
among system variables from measurements of variable configurations or
correlations. As an instance, in the optical waves framework, this
means quantitatively inferring the nonlinear interaction strengths
given the wave emissions. Once the theoretical model is designed,
assuming an effective equilibrium (true thermodynamic equilibrium or
stationary conditions), one has to maximize the likelihood functional
with respect to the coupling parameters. The likelihood functional is
defined as the probability of a variable configurations given the
values of the interaction couplings. For large systems it is
numerically intractable but one can resort to the so-called pseudolikelihood
functional defined as the probability of one variable conditional to
all other variables and to the values of the couplings
\cite{Barber12}.\\ Based on pseudolikelihood maximization, we adopt
two methods to determine the interactions: the well known
$\ell_1$-regularization \cite{Ravikumar10,Aurell12}, that we have
improved with a hypothesis testing procedure based on the evaluation
of the eigenvalues of the Fisher information matrix
\cite{Wasserman03}, and the most recent decimation technique
\cite{Decelle14}.  In order to test the methods, we considered both
the phasor and the $XY$-spin models, generating the data by means of
Monte Carlo numerical simulations. Among the simulated networks we
analyze both sparse graphs, in which the number of interacting
quadruplets $N_q$ scales like the number of variables, $N_q \propto
N$, and dense graphs, in which $N_q \propto N^3$ \footnote{This is a
  diluted dense graph: not all quadruplets are present, though their
  number per variable node scales with $N$, unlike in sparse graphs. A
  complete dense graph would contain $O(N^4)$ interacting
  quadruplets}.  We stress that the techniques here reported might be
applied to any wave system with non-linear collective behavior, such
as phase-locking, breathers and synchronization
\cite{Kuramoto75,Antoni95,Acebron05,Gupta14}, including the prototype
Fermi-Pasta-Ulam model \cite{Fermi55}.  Further on, the methodology
can be translated to simpler cases, e.g., discrete variables models
like the $p$-clock model \cite{Ortiz13,Marruzzo15}, in which rotators
only take $p$ discrete values. Properly modifying the mode interaction
these $p$-clock models can, eventually, represent multi-body Potts
models \cite{Potts52}, as well.

\noindent
\textbf{Results}\\ \textbf{Test models.} Our first test model consists
of $N$ phasors $a_k$ with a global constraint $\sum_{k=1}^N |a_k|^2 =
\mbox{const}\times N$, hereafter termed Spherical Model (SM), with
Hamiltonian \cite{Antenucci15b}
\begin{eqnarray}
{\cal H}&=&-\frac{1}{8}\sum^{\rm d.i.}_{jklm} J_{jklm} ~a_j a_k^* a_la_m^* + \mbox{c.c.}
\label{eq:H}
\end{eqnarray}
The $a_k$'s represent, e. g., the
complex amplitudes of the normal modes expansion of the
electromagnetic field \cite{SargentIII78}
\begin{eqnarray}
{\bm \tilde E}(\bm r, t)=\sum_k a_k(t) \bm E_k(\bm r) e^{\imath \omega_k t} +\mbox{c.c.}
\end{eqnarray}
characterizing the light modes in the $\bm E_k(\bm r)$ basis. The
amplitude $a_k(t)$ is the slowly varying coefficient of the normal
mode $\bm E_k$ of frequency $\omega_k$ and varies on time scales much
slower than $\omega_k^{-1}$.  We adopt it as test model because it is
the lowest order of nonlinearity satisfying time reversal symmetry of
light, as occurring, e. g., in centrometric crystals with symmetric
atomic potentials \cite{Boyd02}.  The laser transition can be
represented as a phase transition in statistical mechanical
theory. This turns out to be possible both in ordered multimode
mode-locked lasers
\cite{Gordon02,Gordon03a,Weill05,Gat10,Antenucci15c, Antenucci15d,
  Marruzzo15} and in random lasers
\cite{Angelani06b,Leuzzi09a,Antenucci15a,Antenucci15b}.  Considering
further orders of the interaction does not change the critical
behavior and the onset of the lasing regime, nor the qualitative
features of the laser in the high pumping regime.  We stress that,
simply in order to focus the presentation, also lower order
interactions (pairwise and three body) are not considered here: the
sum with superscript ``d.i.''  in Eq. (\ref{eq:H}) is intended solely
over quadruplets with distinct indices.

According to multimode laser theory
\cite{SargentIII78,Svelto98,Haus00,Boyd02,Gordon02} modes do interact nonlinearly if and
only if their frequencies satisfy a frequency matching
condition \cite{Antenucci15c},
i.e., given any four modes
$j,k,l,m$ of typical line-width $\gamma$, their angular frequencies are such
that
\begin{eqnarray}
|\omega_j-\omega_k+\omega_l-\omega_m| \lesssim \gamma
\label{eq:FMC}
\end{eqnarray}
at least in one permutation of their indices.
\\
\indent
With equipartite magnitudes ($|a_k|\simeq 1$, $\forall~k$) or with quenched
ones ($|a_k(t)|=A_k(0)$) Eq. (\ref{eq:H}) for phasors reduces to the
so-called $XY$ model for rotators, $a_j=A_j e^{\imath \phi_j} \to
e^{\imath \phi_j}$, with Hamiltonian
\begin{eqnarray}
{\cal H}&=&-\frac{1}{8}\sum^{\rm d.i.}_{jklm} \bigl[J^R_{jklm} ~\cos(\phi_j-\phi_k+\phi_l-\phi_m)
\label{eq:HXY}
\\
&&
\quad+\quad
 J^I_{jklm} ~\sin(\phi_j-\phi_k+\phi_l-\phi_m)\bigr]
\nonumber
\end{eqnarray}
where $J^{R,I}$ are, respectively, real and imaginary parts of the
coupling constants. The $4XY$ model is our second test model. Besides
being an approximation of the SM model, having locally constrained
variables allows for testing the inference techniques also on sparse
graphs at low temperature \footnote{Indeed, it can be seen that for
  bond-disordered SM's, if the node connectivity does not increase at
  least with with $N^2$, all the power $\sum_{k=1}^N |a_k|^2$
  condensates into one single quadruplet below threshold
  \cite{Antenucci15d}.}.  Furthermore, terming $\delta \omega$ the
frequency spacing among the modes, we considered both strict frequency
matching conditions, cf. Eq. (\ref{eq:FMC}), based on comb-like
\cite{Hugi12} single mode resonance distributions ($\gamma \ll
\delta\omega$), as well as {\em narrow-band} conditions ($\gamma >
\delta\omega$). In the latter case Eq. \eqref{eq:FMC} does not play
any role and the node frequencies have no influence on the structure
of the graphs. On the other hand, in graphs built considering $\gamma
\ll \delta\omega$ frequencies do play an important role. These will be
called Mode-Locked (ML) graphs.

We infer data within the Boltzmann-Gibbs equilibrium hypothesis (see
Methods for data generation).  Then, the probability of a
configuration $\bm a$, given a set ${\bm J}$, i.e., the likelihood
functional, reads:
\begin{equation}
\label{eq:p_4a}
P(\bm a|\bm J) = \frac{1}{Z[\bm J]} 
\exp\left\{-\beta{\cal H}[\bm a|\bm J]\right\}
\end{equation}
Computing 
   $Z[{\bm J}]$ is very hard in general. To circumvent this bottleneck one first defines
the single variable pseudo-likelihood \cite{Barber12} of the
values of $a_i$ biased by all other ${\bm a}_{\backslash i}$ values, and by ${\bm J}$
(see details in Methods)
\begin{eqnarray}
P_i(a_i | \bm a_{\backslash i},\bm J) &=& \frac{1}{Z_i[\bm a_{\backslash i},\bm J ]}
\exp\left\{a_i H_i [\bm a_{\backslash i},\bm J] + \mbox{c.c.}\right\}
 \label{eq:p_4ai}
 \end{eqnarray}
 with 
 \begin{eqnarray}
\label{eq:Heff}
H_j[\bm a_{\backslash j},\bm J] &=& \sum^{\rm d.i.}_{klm\neq j}J_{jklm}  {\cal F}_{klm}
\\
\label{eq:F}
{\cal F}_{klm}  &=&\frac{1}{3} \left[ a_k^*a_la_m^* + a_k a_l^*a_m^* + a_k^*a_l^*a_m \right]
\\
Z_i[{\bm a_{\backslash i}},\bm J ]&\equiv& \sum_{a_i} \exp\left\{a_i  H_i[\bm a_{\backslash i},\bm J]+\mbox{c.c.}\right\}
\end{eqnarray}
Further on, if one considers $M$ independent configurations
$\{{\bm a}^{(\mu)}\}$, with $\mu=1,\ldots,M$, the pseudolikelihood
of all the single node variables $\{a^{(\mu)}_i\}$, given all the
others $\{{\bm a}^{\mu}_{\backslash i}\}$ and the couplings ${\bm
  J}$, factorizes. In order to deal with sums instead of products, one
usually evaluates the log-pseudolikelihood that, thus, reads
\begin{eqnarray}
\label{eq:pl_4a}
{\cal  L}^{(0)}_i(\{{\bm a}^{\mu}_{\backslash i}\}, \bm J) &=& 
\sum_{\mu=1}^M  \left(a_i^\mu H_i[\bm a_{\backslash i}^\mu,\bm J] + \mbox{c.c.}\right)\\
\nonumber
&&\quad\quad -\sum_{\mu=1}^M \ln Z_i[\bm a^\mu_{\backslash i},\bm J]
\end{eqnarray}
The $\bm J$'s  maximizing  ${\cal  L}^{(0)}_i$s are considered as the most probable couplings that originate the $\{{\bm a}^{(\mu)}\}$ configurations. 

  We analyze data from systems whose coupling values are randomly
  generated with a bimodal distribution $P(J)=1/2[\delta(J-\hat
    J)+\delta(J+\hat J)] $, where $\hat J=1/N^{(z-1)/2}$, when the
  total number of quadruplets scales as $N_q \sim N^z$.  This is the
  case, e. g., of the frustrated glassy random lasers
  \cite{Antenucci15a,Antenucci15f,Marruzzo16,Ghofraniha15,Gomes15,Pincheira15},
  but the methods here exposed also work for the simpler cases of
  uniform couplings, like in standard mode-locking lasers
  \cite{Haus00,Gordon02,Marruzzo15,Antenucci15c} and random couplings
  with a relative small fraction of negative values, e.g., random
  unfrustrated lasers \cite{Leuzzi09a,El-Dardiry10,Marruzzo15}.

\noindent
\textbf{Data Analysis}\\ \textbf{Decimation and
  $\ell_1$-regularization.} We compare inference predictions obtained
by different implementations of PLM. The $\ell_1$-regularization
consists in adding to Eq. (\ref{eq:pl_4a}) a regularizing term ${\cal
  L}^{(0)}_i - \lambda || J ||_1$, penalizing large $\bm J$
values and it is known to be particularly useful in retrieving
sparse systems\cite{Aurell12}.  The {\em decimation} procedure, instead,
iteratively removes the smallest couplings (cf., Methods).  In this
procedure one maximizes the {\em total} log-pseudolikelihood, summed over
all the modes, i.e.,
\begin{eqnarray}
{\cal L}({\bm J})\equiv {\cal L}= \frac{1}{N}\sum_{i}^N{\cal L}^{(0)}_i
\label{eq:Ltot}
\end{eqnarray} 
It is important to underline that, by maximizing each ${\cal
  L}^{(0)}_i$ separately, cf. Eq. \eqref{eq:pl_4a}, each coupling
$J_{ijkl}$ turns out to be inferred four times with, generally, four
different estimates. The mean value is, then, usually taken as best
reconstructed value. By maximizing the total ${\cal L}$, instead, each
coupling $J_{ijkl}$ is inferred only once.

\noindent
\textbf{Data size and external tuning.}  We consider the effects of
varying the size $M$ of data sets, as well as, the {\em
  temperature}-like parameter $T$ that determines the strength of the
interaction. $T$ resembles real data noise \cite{Tyagi16} or it is
used to drive the system to a phase transition, if present. As it will
be shown, functioning of PLM's qualitatively change at criticality and
in different thermodynamic phases.

\noindent
\textbf{Quality indicators.}  To evaluate the performances of the
techniques we will consider the following quality indicators: (i) the
True Positive Rate (TPR), that is the fraction of true bonds also
appearing in the inferred set of bonds, (ii) the True Negative Rate
(TNR), that is the fraction of missing bonds also absent in the
inferred set of bonds, and (iii) the reconstruction error
 \begin{eqnarray}
 \mbox{err}_J \equiv \sqrt{\frac{\sum_{q}  (J_q-J^*_q)^2}{\sum_q J_q^2}}
 \label{eq:errJ}
 \end{eqnarray}
 yielding how far the inferred values $J_q^*$ of the distinct
 quadruplets $q\equiv \{i,j,k,l\}$ are from the true values $J_q$.
 Exclusively for the decimation PLM, in order to reconstruct the
 number of non-zero couplings, i.e., the number of quadruplets
 actually present in the system, we analyze also the behavior of the {\em tilted} pseudolikelihood function defined
 as:
 \begin{equation}\label{eq:l_t}
 {\cal L}_t\equiv{\cal L}(x)-x {\cal L}_{\rm max}-(1-x){\cal L}_{\rm min}
 \end{equation}
 where $x$ is the number of non-decimated, i.e., non-erased,
 couplings. ${\cal L}_{\rm max}$ is the maximum of the total
 log-pseudolikelihood, Eq. \eqref{eq:Ltot}, at the beginning of the
 decimation procedure, when all possible couplings are contemplated,
 while ${\cal L}_{\rm min}$ is evaluated on a graph with no
 links. ${\cal L}(x)$ is the maximum with respect to the $x$ fraction
 of all possible couplings that are still considered to be important
 parameters of the problem.  Erasing irrelevant couplings does not
 affect ${\cal L}(x)$ so that a plateau occurs in $x>x^*$ until
 important couplings start to be decimated and ${\cal L}(x)$ starts to
 decrease. In order to ease the identification of the optimal number
 of fitting parameters $x^*$, ${\cal L}(x)$ is tilted: the optimal
 value $x^*$, corresponding to the amount of couplings in the true
 network, is determined looking at the maximum of ${\cal L}_t$.


     \begin{figure}[t!]
   	\centering
  \includegraphics[width=1\columnwidth]{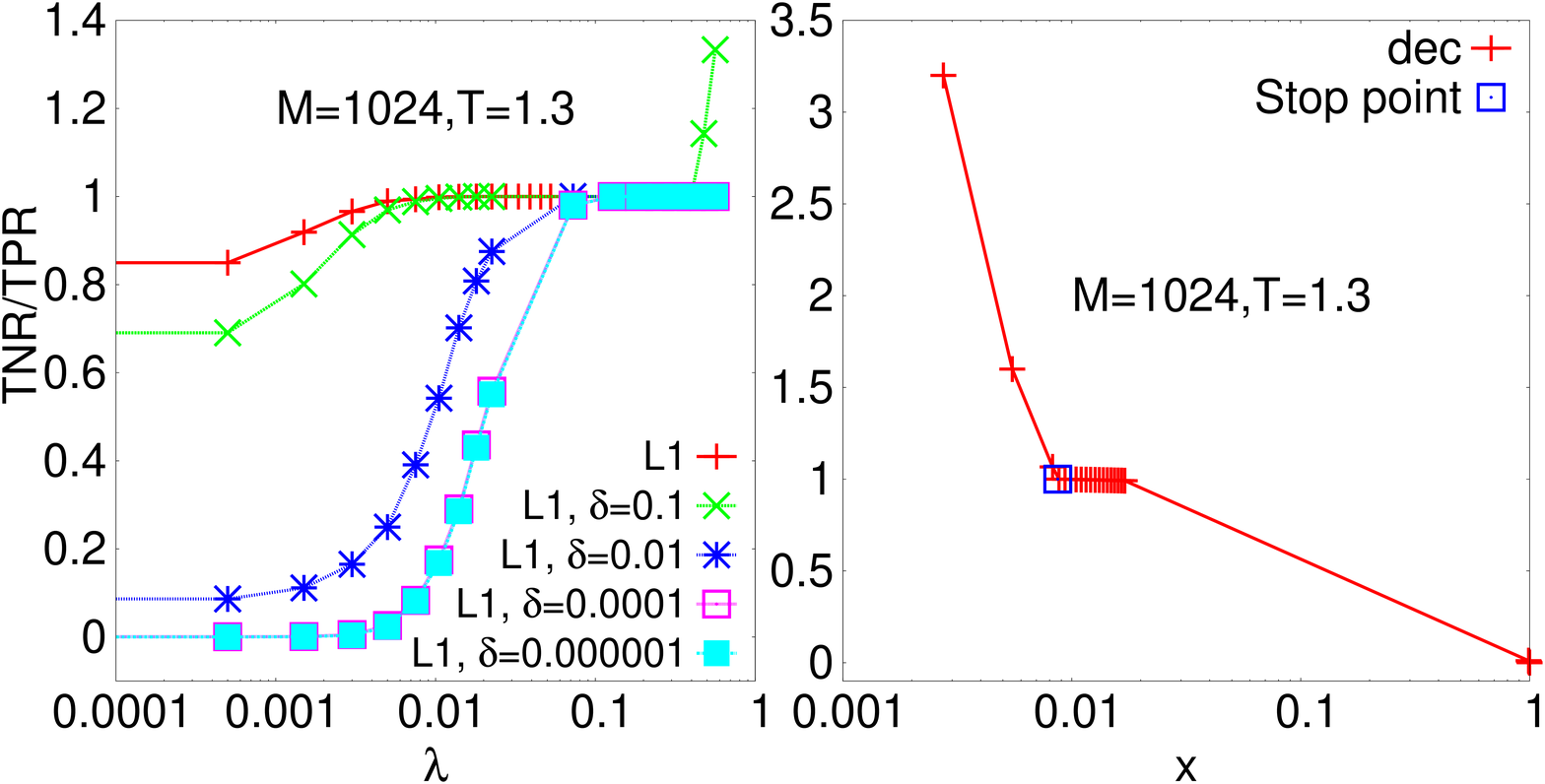}
   	\caption{The TNR/TPR ratio vs. the regularizer $\lambda$ used
          for the $\ell_1$-regularization (left) and vs. the fraction
          $x$ of undecimated couplings for the PLM with
          decimation. The stopping point indicates the maximum of
          $\mathcal{L}_t$, Eq. \eqref{eq:l_t}, where the decimation
          procedure stops. In the first case two different criteria
          are chosen to eliminate small bonds: the {\em a-priori}
          $\delta$ thresholding or the {\em a posteriori} inferred
          bond distribution thresholding based on the Fisher
          information matrix (see Methods for details). Data are
          taken from a $4XY$ model on a sparse Erdos-Renyi random
          graph with $N_q=N$, $N=16$, $T=1.3$, $M=1024$.  In this case
          the finite size proxy for the critical temperature is
          $T_c(16) \simeq 1.34$.}
   	\label{fig:tnr_tpr}
   \end{figure} 
  
In Fig. \ref{fig:tnr_tpr}, using data from a $4XY$ model on
Erdos-Renyi (ER)-like sparse graph, we show how the TNR/TPR ratio
increases to $1$ as the parameter $\lambda$ used for the
$\ell_1$-regularization, i.e., ${\cal L}^{(0)}_i - \lambda || J ||_1$,
is increased. Further on a $\delta$-threshold criterion
\cite{Aurell12} is adopted for {\em model selection}, i.e., the
ability to reduce the number of parameters to the relevant ones.
Within the $\delta$-threshold criterion, couplings which are inferred,
in absolute value, to be less than $\delta$ are considered to be
irrelevant and are set to zero.  The value for $\delta$ is, however,
chosen {\em a priori} and the choice might be delicate when there is
not a clear gap in the distribution of the inferred couplings
\cite{Decelle14}.  Moreover, as $\lambda$ is small, we see that the
smaller the $\delta$ the less precise the network reconstruction. On
the other hand, the smaller the $\lambda$ the less perturbed the
original pseudolikelihood (PL). Indeed, increasing $\lambda$ the
chance of globally underestimate the couplings increases.

\begin{figure}[t!]
   	\centering
   	\includegraphics[width=1\linewidth]{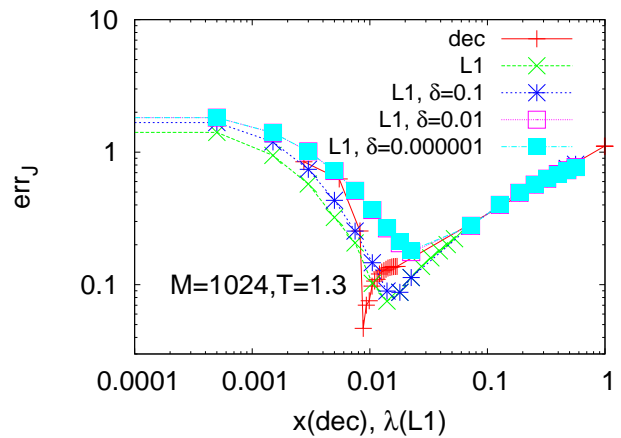}
   	\caption{Reconstruction error for the $4XY$ model on sparse
          Erdos-Renyi graph with $N_q=N=16$, $M=1024$ at $T=1.3$.  The
          error obtained following various $\ell_1$-regularized PLMs
          is displayed vs. $\lambda$; the decimation PLM
          reconstruction error is plotted against the fraction of
          non-decimated couplings $x$.  }
   	\label{fig:err_J}
   \end{figure}

If the probability distributions of the estimators are known, the
issues related to an {\em a priori} fixing of a $\delta$ threshold
might be overcome through a more accurate {\em hypothesis testing}
procedure. Indeed, it can be seen that, as $M \rightarrow \infty$, the
probability distribution of the maximum PL estimator is a Gaussian
with variance given by the diagonal element of the inverse of the
Fisher information matrix \cite{Wasserman03}. Therefore, as detailed
in Methods, we can construct a confidence interval for each
estimated value and verify whether it is compatible with the
hypothesis ``being a zero coupling''. If it is the case, it is
considered as an irrelevant parameter and erased. As we can see from
Figs. \ref{fig:tnr_tpr} and \ref{fig:err_J} this criterion for model
selection outperforms, for every value of $\lambda$, the
$\delta$-threshold method. Moreover, as detailed in the Methods,
this criterion provides a method to determine the best value for the
regularizer $\lambda$ that is usually chosen arbitrarily.

 Always in Fig.  \ref{fig:tnr_tpr} (right) we display the TNR/TPR
 ratio obtained with the decimation PLM as the fraction of
 non-decimated couplings $x$ decreases (from fully connected limit
 $x=1$ to non-interacting graph $x=0$).  At $x=1$ the TPR is always
 one, for any $M$ and $T$, whereas the TNR$=0$.  As the fraction of
 non-decimated couplings decreases but remains greater than or equal
 to the true one ($x^*=2/15$ in the original model analyzed in the
 right panel of Fig.  \ref{fig:tnr_tpr}) the TPR does not decrease and
 the TNR increases towards one. Eventually, more couplings than those
 of the original network are decimated: the TPR starts decreasing and
 the ratio TNR/TPR consequently grows above $1$ as $x \to 0$. The blue
 square indicates the stopping point of the decimation procedure
 determined, instead, as the maximum of the $\mathcal L_t$,
 cf. Eq. \eqref{eq:l_t}. It can be observed that in this case it
 perfectly reconstructs the network of interactions since the
 TPR=TNR=$1$.

 \textbf{Decimation PLM.}  When comparing the performances of the most
 efficient regularization method with the decimation one, we observe
 that the network reconstruction in terms of true and false couplings
 is very adequate with both methods.  However, the order of magnitude
 of the reconstruction error, testing also the quality of the inferred
 {\em values} of the couplings, is smaller in the decimation PLM,
 cf. Fig. \ref{fig:err_J}, when the fraction of decimated couplings
 ($1-x$) equals the one of the true network.  It is important to
 underline that, within the decimation PLM, no parameters are
 determined {\em a priori}: the optimum value of $x$ is determined
 maximizing the $\mathcal L_t$. Moreover, exact fraction of relevant
 parameters and best estimate of their values are simultaneously
 inferred, which is not always true in the PLM with
 $\ell_1$-regularization since even the smallest $\lambda$ for network
 reconstruction might induce a too high global underestimation of the
 couplings.  We, thus, deepen the analysis of the decimation PLM.
       
   \begin{figure}[t!]
   	\centering
 	\includegraphics[width=1\columnwidth]{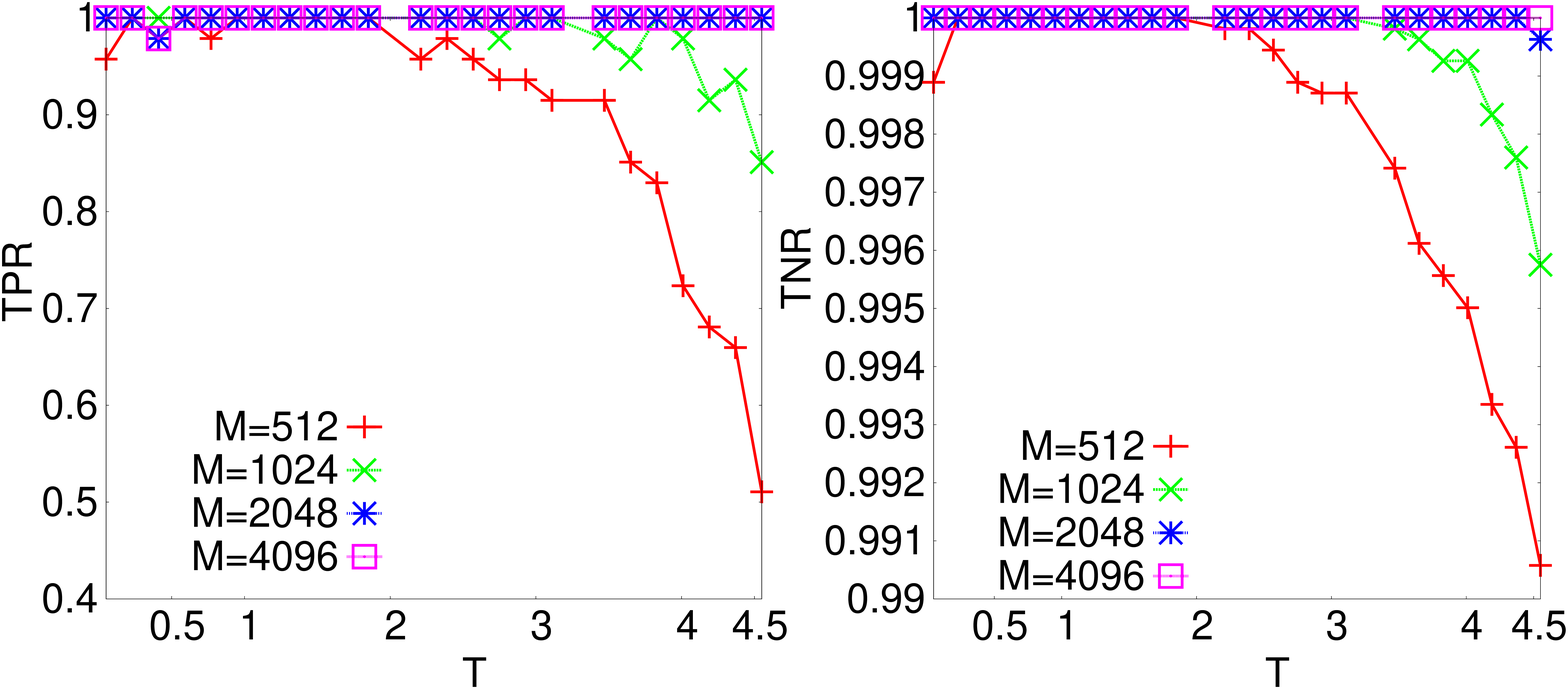}
   	\caption{ TPR (Left) and TNR (Right) for decimated networks at
          the maximum $x^M$ of the tilted pseudolikelihood $\mathcal
          L_t$ vs $T$ at different data-set sizes $M$ for the $4XY$
          model on sparse Mode-Locked graphs with $N=16$, $N_q=47$,
          $T_c(N)\simeq 0.50$ ($T_c(\infty)=0$).  }
   	\label{fig:ROC_M_T}
   \end{figure} 

In Fig. \ref{fig:ROC_M_T} we display TPR (left) and TNR (right)
vs. $T$ and $M$ for the decimated network at the maximum point, $x^M$,
of the $\mathcal L_t$ for the $4XY$ model on a Mode-Locked-like sparse
graph with $N_q=47$ number of quadruplets and $N=16$ nodes. For large
enough $M$ the reconstruction is optimal for all temperatures, whereas
for small $M$ it is guaranteed only in a $T$ interval around the {\em
  finite size } proxy to the critical temperature (see
Methods). Indeed, we observed that, tuning the external
temperature-like parameter for each system studied, one can identify a
``critical'' $T$ interval where the reconstruction error is minimal,
even orders of magnitude smaller than outside such interval, and, more
in general, the system is better and easier reconstructed.  In
Fig. \ref{fig:tplf}, the behavior of the $\mathcal L_t$ versus $x$ is
compared to err$_J(x)$ for three different systems, the $4XY$-model on
ML and ER sparse graphs and the $4$SM-model on a dense ML graph.  All
systems have $N=32$ variable nodes while the number of interaction
quadruplets is $N_q=32,72$ and $2360$, respectively. The number of
configurations in all cases is $M=65000$.  It is clearly observed
that, given a large enough $M$ and/or a critical-like $T$ the maximum
point of the $\mathcal L_t$, $x^M$, coincides with the minimum point
of err$_J$, $x^m$.  The decimation PLM gives then a criterion to
determine the number of interaction couplings in the system from
measurements data without any {\em a priori} chosen parameters.  We
notice that, as $M$ is small and $T$ far from the critical region, the
maximum point of $\mathcal L_t$ and minimum point of err$_J$ can be
mismatched, as shown in Fig. \ref{fig:x_x}. In
Fig. \ref{fig:err_J_T_M} the $T$ dependence of err$_J$ is plotted for
the $4XY$ model on a ER sparse graph and for the $4$SM model on a ML
dense graph at different values of $M$.  As detailed in Methods the critical $T$ interval turns out to be identified by the
(finite size) critical temperature estimate of the phase transition
point of the direct statistical mechanical problem.\\

        
        \begin{widetext}
            \begin{center}
     \begin{figure}[h!]
  		\includegraphics[width=.8\textwidth]{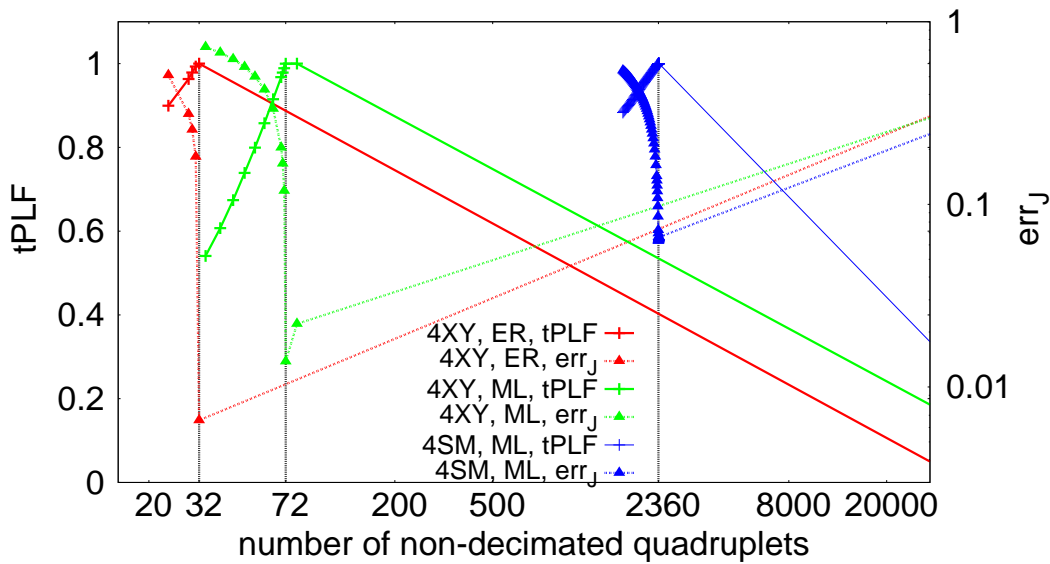}
   	 \caption{ Tilted pseudolikelihood tPLF (normalized to its
           maximum) and reconstruction error for different
           models. Concave curves are tPLF, convex
           curves are err$_J$. Three apart models on different
           random graphs are considered. Red (left) curves: $4XY$
           model on sparse Erdos-Renyi graph with $N=32, N_q=32$,
           $M=65000$, at $T=1.2$ ($T_c(32)\simeq 1.39$). Green (mid)
           curves: $4XY$ model on sparse Mode-Locked graph with $N=32,
           N_q=72$, $M=65000$, at $T=1.8$ ($T_c(32)\simeq 0.72$). Blue
           (right) curves: $4$SM model on dense Mode-Locked graph
           with $N=32, N_q=2360$, $M=65000$, at $T=6.2$
           ($T_c(32)\simeq 0.91$).  }
   	\label{fig:tplf}
	   \end{figure} 
	   \end{center}
    \end{widetext}

     \begin{figure}[t!]
   	\centering
   	\includegraphics[width=1\linewidth]{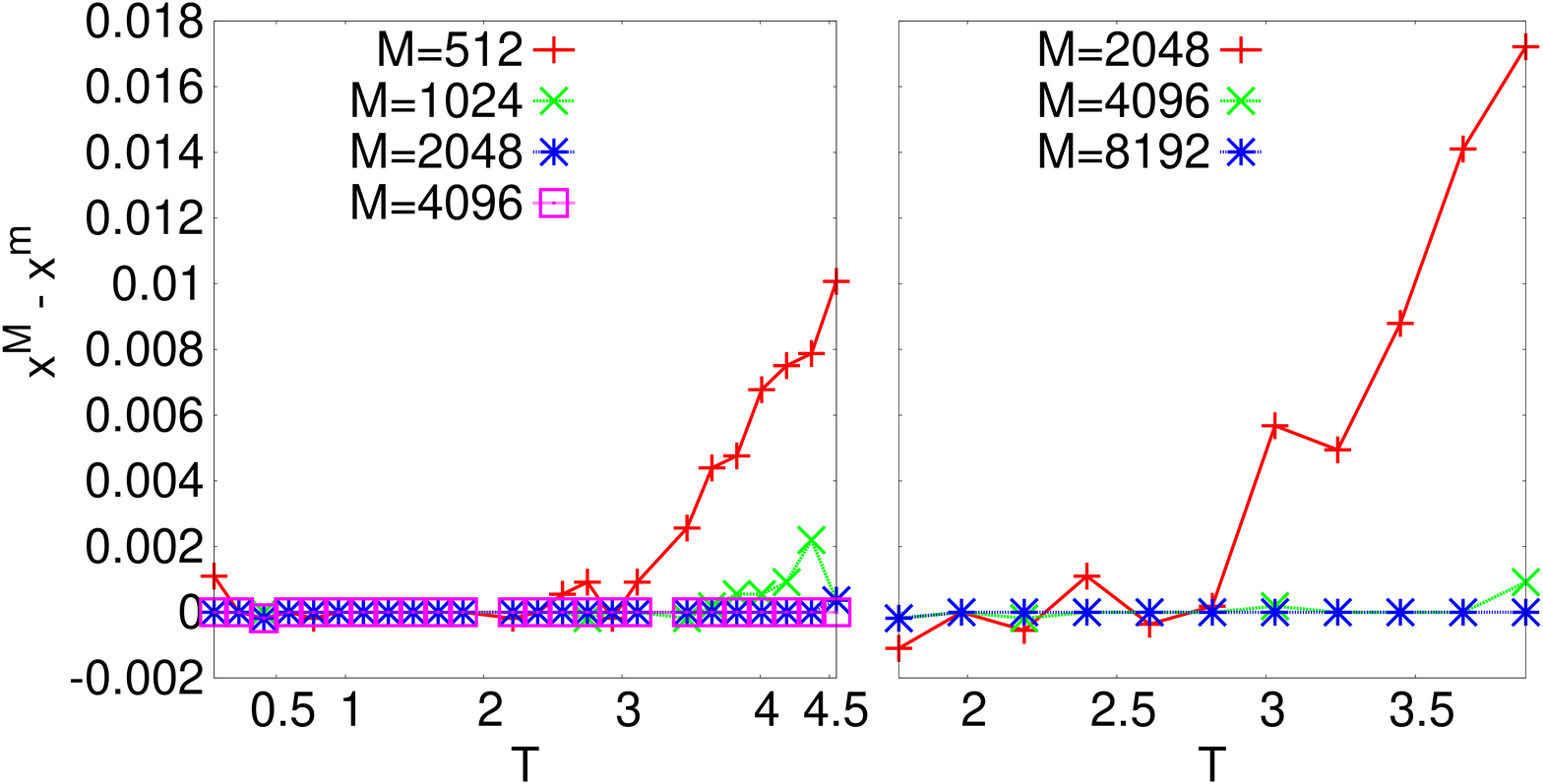}
   	\caption{Plot of the difference between the maximum point of
          $\mathcal L_t$, $x^{\rm M}$, and the minimum point of the
          reconstruction error, $x^{\rm m}$, vs $T$ at different $M$
          for systems of $N=16$ variables. Left: $4$-$XY$ model on
          sparse Mode-locked graph with $N_q=47$ ($T_c(16)\simeq
          0.50$). Right: $4$SM model on dense Mode-Locked graph
          with $N_q=252$ ($T_c\simeq 1.07$).  }
   	\label{fig:x_x}
   \end{figure} 
       
     \begin{figure}[t!]
   	\centering
	\includegraphics[width=1\linewidth]{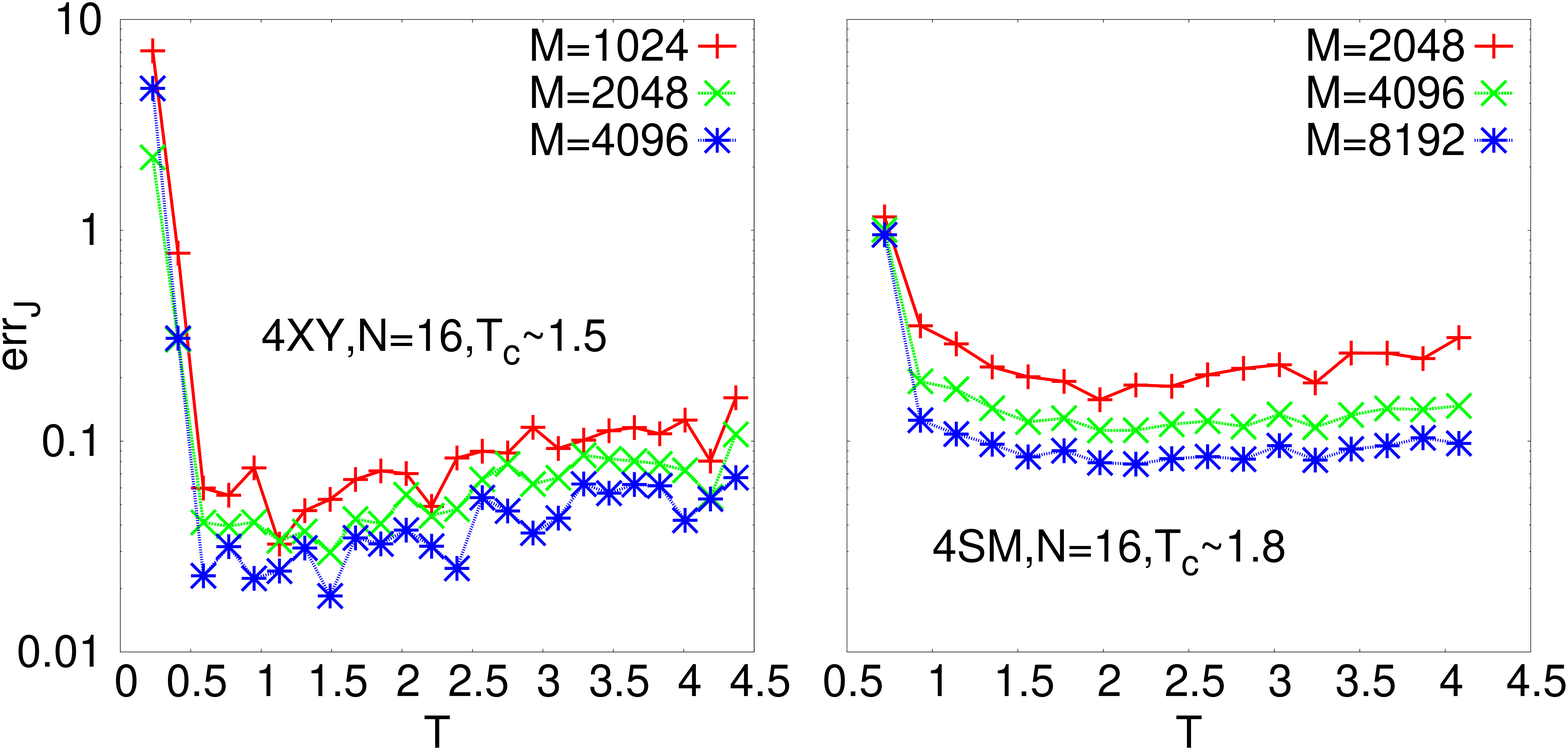}
   	\caption{Reconstruction error at its minimum $x^{\rm
            min}=x^{\rm true}$ in the decimation procedure vs. $T$ for
          various $M$ for $N=16$ variables systems with bimodal random
          values of the coupling constants.  Left: $4XY$ model on
          sparse Erdos-Renyi graph with $N=16, N_q=16$. Right: $4$SM model on dense Mode-Locked graph
          with $N=16, N_q=252$. }
   	\label{fig:err_J_T_M}
   \end{figure}

\noindent
\textbf{Discussion}\\
\noindent
We have been applying and improving Pseudo-Likelihood Maximization(PLM) techniques to the
inverse problem in multi-body models, representing systems with
nonlinerar response in generic theories. 
Firstly, we have quantitatively measured the quality of the network reconstruction showing that both PLM methods analyzed allow to obtain an optimal reconstruction for these systems in all thermodynamic phases, when large enough number of samples $M$ are available. Decreasing $M$ the optimal reconstruction is better achieved around the finite size proxy to the critical temperature. Performing then a deeper analysis of the reconstruction error, which gives information on how far the inferred couplings are from the true couplings, reveals that with the decimation PLM the inferred values are closer to those ones of the original systems.    
Our analysis has been
motivated by the study of lasers in the framework of statistical
mechanics, though, looking at the models employed, Eqs. (\ref{eq:H},\ref{eq:HXY}), its range of applicability is more
widespread and potentially involves many problems in which both
nonlinear and multi-body contributions turn out to be relevant in
determining the system behavior, see, e.g.,
\cite{Haus00,Wiersma08,Mezard02,MacKay03,Mezard09,Goetze09,Katz11}.
Focusing on optics, data from experiments would allow to identify
active and passive mode-locking in multimode lasers in terms of
mode-coupling coefficients. When more modes on the network graph are
connected by a non-vanishing coupling they are matched in frequency,
cf. Eq. (\ref{eq:FMC}), and therefore, beyond some critical point,
they will be locked in phase. Configurations of magnitudes and phases
can be obtained from the Fourier analysis of the pulses in ultrafast
multimode lasers \cite{Trebino93,Trebino97,Trebino02}, pulses known to
occur because of mode-locking, and parameters like the self amplitude
modulation coefficients of saturable absorbers and the Kerr parameter
can be inferred.  In presence of relevant light scattering, instead,
occurring in random lasers, no direct measurements of phases has been
carried out so far, to our knowledge, but only spectral intensities,
i.e., modes magnitudes. Acquiring also phases configurations, our
inference technique would give the possibility to determine the
strength of the interaction among the modes in the systems and to
discriminate whether or not self-starting mode-locking occurs in
random lasers.

\noindent
\textbf{Methods}\\
\noindent
\textbf{Data Generation.}~ Configurations are generated by means of
Monte Carlo numerical simulations of the Hamiltonians of the SM,
Eq. (\ref{eq:H}), and the XY, Eq. (\ref{eq:HXY}) models with number of
modes $N=8,16,32$.  In the SM model diluted dense graphs are
simulated, i. e., with a total number of quadruplets $N_q \sim N^3$
and with an average number of couplings per variable $N_q/N=
O(N^2)$. We generate graphs of two kinds. One kind is {\em \`a la}
Erdos-Renyi: every mode participates in a number of couplings chosen
with a Poissonian distribution but independently from each other. The
other kind is Mode-Locked: a mode interacts with other three modes
only if their frequencies satisfy one of the frequency matching
condition permutations, cf. Eq. (\ref{eq:FMC}).  In the XY model
graphs are sparse, i. e., $N_q/N= O(1)$ and we generate both models on
standard ER graphs and on ML graphs.  In order to speed up
thermalization and have access to data at equilibrium the Parallel
Tempering algorithm has been used. Configuration recording is
performed at Monte Carlo steps far enough to avoid data correlation in
the critical region.

\noindent
\textbf{Maximization algorithm.}~ In our analysis, rather than
maximizing ${\cal L}^{(0)}_i({\bm J}) - \lambda || J ||_1$ and ${\cal
  L}(x,{\bm J})$, we invert the functional signs and minimize, in
order to use well established packages available. For the minimization
of -PL with $\ell_1$-regularization we adopted L1General
\cite{schmidt10}: a set of MATLAB routines implementing several of the
available unconstrained minimization strategies for solving
$\ell_1$-regularization problems. In particular, the results here
presented were obtained with the spectral projected gradient method
\cite{schmidt10}.  For the minimization of $-{\cal L}(x,{\bm J})$ in
the decimation procedure, we use, instead, Open Optimization Library
(OOL) \cite{ool}, which is a set of constrained optimization codes
written in C. The results here presented were obtained using the
spectral projected gradient method for convex set \cite{Birgin00}. The
decimation code using the OOL routine has been {\em ad hoc} developed
with parallel programming on GPU's, reducing the running times of
numerical processing with respect to C++ codes on CPU up to a factor
$5$ for the largest analyzed systems, $N=32$. In terms of timing, the
parallel CUDA code outperforms the MATLAB routines up to a factor $30$
for $N=16$.

\textbf{Inverse algorithm} To evaluate the inverse of the Fisher
Information matrix, defined by
\begin{equation}
\mathcal{I}^i_{a b} = - \frac{\partial^2 \mathcal{L}_i}{\partial J_a \partial J_b}\bigg|_{\hat{\mathbf{J}}}
\end{equation}
where with $a$, $b$ we indicate two possible quadruplets node $i$
might belong to (see next paragraph), we first evaluate its Cholesky
decomposition, being $\mathcal{I}^i_{a b}$ a symmetric, positive
define square matrix, and then its inverse through GSL libraries.\\
\textbf{Pseudolikelihood function} From Eq. (\ref{eq:H}) in the main text, the likelihood function of a mode amplitude
configuration $\bm a$, given a coupling set ${\bm J}$, is readily
introduced as
\begin{equation}
\nonumber
P(\bm a|\bm J) = \frac{1}{Z[\bm J]} 
\exp\left\{-\beta{\cal H}[\bm a|\bm J]\right\}
\end{equation}
To construct the likelihood function of $a_j$, i. e., the probability
distribution of a single variable given the values of all the other
variables $\bm a_{\backslash j}$, we first rewrite Eq. (\ref{eq:H}) in
an equivalent way:
\begin{eqnarray}
\nonumber
{\cal H}[\bm a|\bm J]&=& -\frac{1}{8}\sum_{j=1}^N a_j  H_j[\bm a_{\backslash j},\bm J] + \mbox{c.c.}
\end{eqnarray}
defining the complex-valued local effective fields
\begin{eqnarray}
\nonumber
H_j[\bm a_{\backslash j},\bm J] &=& \frac{1}{4}\sum^{\rm d.i.}_{klm\neq j}J_{jklm}  {\cal F}_{klm}
\\
\nonumber
{\cal F}_{klm}  &=&\frac{1}{3} \left[ a_k^*a_la_m^* + a_k a_l^*a_m^* + a_k^*a_l^*a_m \right]
\end{eqnarray}
Then, we separate the contributions from a given variable $a_i$ from
all contributions not involving $a_i$, only $\bm a_{\backslash i}$:
\begin{eqnarray}\nonumber
{\cal H}[\bm a|\bm J] &=& -\frac{1}{8}a_i H_i[\bm a_{\backslash i},\bm J] \\ \nonumber
 & &-\frac{1}{8}\sum_{j\neq i}^{1,N} a_j  H_j[\bm a_{\backslash j},\bm J]  + \mbox{c.c.}
\label{eq:H_Hi}
\\
&=& {\cal H}_i[a_{i}|\bm a_{\backslash i},\bm J] + {\cal H}_{\backslash i}[\bm a_{\backslash i}|\bm J]
\end{eqnarray}
In terms of this decoupling the partition function reads:
\begin{eqnarray}
Z &=&
 \sum_{ \bm a_{\backslash i}} \exp\left\{-\beta {\cal H}_{\backslash i}[\bm a_{\backslash i}|\bm J]  \right\}  Z_i[{\bm a_{\backslash i}},\bm J ]
 \label{eq:Z_Zi}
\end{eqnarray}
with 
\begin{eqnarray}
\nonumber
Z_i[{\bm a_{\backslash i}} ]\equiv \sum_{a_i} \exp\left\{a_i  H_i[\bm a_{\backslash i},\bm J]+\mbox{c.c.}\right\} , 
\end{eqnarray}
i. e., Eq. (9) of the main text.
In order to effectively carry out the sum over $a_i$ values one has to recall that the phasors satisfy a global {\em spherical} constraint $\sum_j |a_j|^2 = \epsilon N$, with constant $\epsilon$.
Once all $\bm a_{\backslash i}$ are given, then, the value of $|a_i|$ is fixed, 
\begin{eqnarray}
\nonumber
|a_i| = \sqrt{\epsilon N - \sum_{j \backslash i} |a_j|^2} 
\end{eqnarray}
and $\sum_{a_i}$ simply reduces to an integral on the angular phase variable $\phi_i\in[0:2\pi[$.
Using Eqs. (\ref{eq:H_Hi}) and (\ref{eq:Z_Zi}) of this section the pseudolikelihood  function of the values of  $a_i$, biased by $\bm a_{\backslash i}$ values, can be, eventually, written as
\begin{eqnarray}
P_i(a_i | \bm a_{\backslash i},\bm J) &=& \frac{1}{Z_i[\bm a_{\backslash i},\bm J ]}
\exp\left\{a_i H_i [\bm a_{\backslash i},\bm J] + \mbox{c.c.}\right\}
\nonumber
\end{eqnarray}
 In order to find the best estimates of the 
interaction parameters given a data set of $M$ mode amplitude configurations, we  minimize the opposite
 log-pseudolikelihood functional $-{\cal L}_i$:
\begin{eqnarray}
-{\cal  L}_i &=& 
-\sum_{\mu=1}^M  \left(a_i^\mu H_i[\bm a_{\backslash i}^\mu,\bm J] + \mbox{c.c.}\right)
+\sum_{\mu=1}^M \ln Z_i[\bm a^\mu_{\backslash i},\bm J]
\nonumber
\\
\label{eq:pl_4a_APP}
\end{eqnarray}
with respect to the coupling parameters, 
exploiting the explicit knowledge of the derivatives
\begin{eqnarray}
\label{eq:Dpl_4a}
\frac{\partial (- {\cal L}_i)}{\partial J_{ijkl}} = \sum_{\mu=1}^M {\cal F}^\mu_{jkl}\left[\langle a_i \rangle_i^\mu -
a_i^\mu\right]
\end{eqnarray}
where we denoted
\begin{eqnarray}
\langle \left(\ldots \right) \rangle_i \equiv \frac{1}{Z_i[\bm a^\mu_{\backslash i},\bm J]}
\sum_{a_i} \left( \ldots \right) \exp\left\{
a_i H_i[\bm a_{\backslash i}^\mu, \bm J]+ \mbox{c.c.}
\right\} .
\nonumber
\end{eqnarray}

Rewriting the complex amplitude in polar coordinates $a_i=A_i e^{\imath \phi_i}$
we have the following expression for the marginal (we do not write the $\bm J$ dependence)
\begin{eqnarray}
P_i(A_i, \phi_i | \bm A_{\backslash i}, \bm \phi_{\backslash i}) &=& \frac{\exp\left\{
A_i\left[H_i^R~\cos\phi_i +H_i^I ~\sin\phi_i \right]
\right\}
}{Z_i[\bm A_{\backslash i}, \bm \phi_{\backslash i} ]}
\nonumber
\\
&=& \frac{\exp\left\{ A_i |H_i| \cos(\phi_i - \gamma_i)\right\}}{2 \pi \int dA_i ~I_0(A_i|H_i|)}
\label{eq:p_4xy}
\end{eqnarray}
where
\begin{eqnarray}
\nonumber
|H_i|&=&\sqrt{\bigl(H_i^R\bigr)^2+\bigl(H_i^I\bigr)^2}
\\
\nonumber
\gamma_i&=& \arctan \frac{H_i^I}{H_i^R}
\end{eqnarray}
and $I_0(x)$ is the modified Bessel function of the first kind. 

As mentioned in the main text, the polar coordinates are most
useful in the cases of intensity equidistribution among the modes,
$A_i\simeq 1$, $\forall ~i$, and of quenched amplitudes, i.e., when
the $A_i$ dynamics is quenched on the time scales of the $\phi$'s
dynamics. In the latter case all $A$'s are taken care of by
rescaling the coupling constants as $A_iA_jA_kA_l J_{ijkl}\to
J_{ijkl}$.  When the couplings are considered real-valued, the polar
expressions of the local effective fields, cf. Eq. (\ref{eq:Heff})  in the main text,
can
be rewritten by substituting Eq. (\ref{eq:F}) with
\begin{eqnarray}
{\cal F}^R_{jkl}& = & \cos \phi_j \cos \phi_k \cos \phi_l 
+ \frac{1}{3}(\cos \phi_j \sin \phi_l \sin \phi_k \nonumber \\
&+& \cos \phi_l \sin \phi_j \sin \phi_k + \cos \phi_k \sin \phi_j \sin \phi_l )
\quad
\nonumber
 \\
{\cal F}^I_{jkl} & = &  \sin \phi_j \sin \phi_k \sin \phi_l + \frac{1}{3}(\sin \phi_j \cos \phi_l \cos \phi_k \nonumber \\
&+& \sin \phi_l \cos \phi_j \cos \phi_k + \sin \phi_k \cos \phi_j \cos \phi_l ) \quad
\nonumber
\end{eqnarray}
and the log-pseudolikelihood functional ${\cal L}_i$ and its gradient,
cf. Eqs. (\ref{eq:pl_4a_APP},\ref{eq:Dpl_4a}) in this section, simplify to
\begin{eqnarray}
\label{eq:pl_4xy}
-{\cal L}_i &=& \sum_{\mu=1}^M 
\Biggl\{
\ln 2 \pi I_0 \left(
\left| H_i (\bm \phi_{\backslash i}^\mu) \right|  \right)
\\
\nonumber
&& - \Bigl[ H^R_i(\bm \phi_{\backslash i}^\mu)  \cos \phi_i^{\mu} 
+H_i^I(\bm \phi_{\backslash i}^\mu) 
 \sin \phi_i^{\mu}  \Bigr]
\Biggr\}
\\
\nonumber
 \frac{\partial (-{\cal L}_i)}{\partial J_{ijkl}} &=& \sum_{\mu=1}^M
 \Biggl\{
   {\cal F}_{jkl}^\mu \\
   &&\times \left[\frac{I_1(|H_i(\bm \phi_{\backslash i}^\mu)| )}{I_0(|H_i(\bm \phi_{\backslash i}^\mu) |)}\frac{H_i(\bm \phi_{\backslash i}^\mu) }{|H_i(\bm \phi_{\backslash i}^\mu) |} - e^{\imath \phi_i^\mu}\right] 
   \nonumber
 + \mbox{c.c.} \Biggr\} 
   \end{eqnarray}

To determine the
interaction network of non-linear wave systems we use and compare two techniques:
 the $\ell_1$-regularization PLM \cite{Ravikumar10,Aurell12} and the
decimation PLM \cite{Decelle14}.\\ 
 \textbf{$\ell_1$-regularization} In this approach we add a $\ell_1$ norm
contribution for each coupling to be inferred to the log-pseudolikelihood in
order to keep the coupling values from diverging during the minimization procedure:
\begin{eqnarray}\label{eq:l_1_norm}
-{\cal L}_i \to -{\cal L}_i +\lambda \sum_{jklm}^{\rm d.i.} |J_{jklm}|
\end{eqnarray}

The positive regularizer $\lambda$ must be small in order to prevent the
modification of the landscape of ${\cal L}_i$. We take a pseudolikelihood that is intensive in $M$ but 
small enough values of $\lambda$ for all $M$'s.

Within this method, for each mode $i$ all couplings involving $i$ are
inferred in one apart iteration.  The same quadruplet $J_{ijkl}$ can, thus, be
inferred more times, proceeding by minimizing likelihood functions for
different, though interacting, modes $i$, $j$, $k$ and $l$. Nothing prevents the values
of apart reconstructions to be the same. Eventually, thus, all
inferred values of the same coupling are averaged in order to enforce
the original symmetry.\\
\indent 
\textit{$\delta$ thresholding} A further improvement in the reconstruction of the topology can be achieved, as suggested in Ref. \cite{Aurell12}, by setting to zero all couplings whose estimate is below a threshold value $\delta$. However, the choice of the $\delta$ value might be delicate since there are many cases in which there is no clear gap between the zero and the non-zero couplings \cite{Decelle14}.
If the probability distributions of the estimators are known, we can overcome this problem developing a more accurate hypothesis testing scheme.\\
\indent 
\textit{$\mathcal{P}(\hat{J})$ thresholding} It can be seen that, as $M \rightarrow \infty$, the probability distribution $\mathcal P(\hat{J})$ of the maximum  PL estimator $\hat{J}$ converges to a Gaussian distribution centered around the true value of the coupling and with variance  estimated   by the diagonal elements of the inverse of the Fisher information matrix~\cite{Wasserman03}. The elements, $\mathcal{I}^i_{a b}$,  of the Fisher information matrix are defined through:
    \begin{equation}\label{eq:definition_hessian}
    \nonumber
    \mathcal{I}^i_{a b} = - \frac{\partial^2 \mathcal{L}_i}{\partial J_a \partial J_b}\bigg|_{\hat{\mathbf{J}}}
    \end{equation}
    where with $a$, $b$ we indicate two possible quadruplets  including node $i$, i.e., $a=\{i,j,k,l\}$, $b=\{i,j',k',l'\}$.
Then, knowing the distribution, we can determine, for every estimated value $J^*_a$, if it is compatible with a Gaussian centered in zero, i.e., if the hypothesis for the true coupling to be zero might or not be rejected. 

The hypothesis testing can be schematically developed  as follows.
\begin{enumerate}
\item[(i)] Once the maximum points of the PL are found, we evaluate the inverse of the Fisher information matrix, Eq. \eqref{eq:definition_hessian}; from Eq. \eqref{eq:pl_4xy}, we have:
\begin{eqnarray}
    \label{eq:i_4xy}
    \mathcal{I}^i_{a b} &=& \sum_{\mu=1}^M
    \mathcal{F}^\mu_{a}\mathcal{F}^\mu_b \Biggl\{ \\ \nonumber
    && \times \biggl(\frac{H_i(\bm \phi_{\backslash i}^\mu)}{\left| H_i (\bm \phi_{\backslash i}^\mu) \right|}\biggl)^2
   \mathcal{B}\left(\left| H_i (\bm \phi_{\backslash i}^\mu) \right|\right)\\ \nonumber
    && +\frac{I_1(|H_i(\bm \phi_{\backslash i}^\mu)| )}{I_0(|H_i(\bm \phi_{\backslash i}^\mu) |)} \biggl(\frac{\left| H_i (\bm \phi_{\backslash i}^\mu) \right|^2 - \left(H_i(\bm \phi_{\backslash i}^\mu)\right)^2}{\left| H_i (\bm \phi_{\backslash i}^\mu)\right|^3}\biggr)\Biggr\}
    \end{eqnarray}
    where with $\mathcal{B}\left(x\right)$ we indicate:
    $$
    \mathcal{B}\left(x\right) = \frac{1}{2}\left(\frac{I_2(x)}{I_0(x)}+1\right)-\left(\frac{I_1(x)}{I_0(x)}\right)^2
    $$
The diagonal terms of the inverse matrix are taken as estimates for the variances, $\sigma^*_a$, of the estimator distributions.
\item[(ii)] We assume every couplings to be zero;  hence, we expect every estimated value $J^*_a$ to be compatible with 
 $N\left(0, \sigma^*_a\right)$ distribution. We,  then, construct a confidence interval $C_n$ that should contain the estimated value $J^*_a$ within a  $97.5\%$ probability.
\item[(iii)] If the inferred $J^*_a$ is contained in $C_n$ we accept the zero hypothesis and the coupling is set to zero.
\end{enumerate}
\indent 
\textit{Optimal  $\lambda$ regularizer}
It is clear from Eq. \eqref{eq:l_1_norm}, that as $\lambda$ increases it increases the tendency to globally underestimate the interaction couplings. Within the hypothesis testing procedure described more and more couplings  become compatible with a Gaussian centered in zero and are considered as irrelevant. Indeed, as noticed in Fig. 1, $\mbox{err}_J(\lambda)$ increases above some optimal value $\hat{\lambda}$ that should depend on the number of configurations $M$ available and on $T$. One procedure that might be adopted to determine $\hat{\lambda}$ consists in testing the PLM with $\ell_1$-regularization on data relative to a known system and use the optimum value found to solve the inverse problem of new systems. However, there might be cases in which there are no solved inverse problems available and, in general, $\hat{\lambda}$ could depend on the details of the problem.\\
 The hypothesis testing procedure explained in the previous paragraph could be considered also as a tool to determine $\hat{\lambda}$. Indeed, within the PLM, every coupling $J_{ijkl}$ has four estimators $J^*_a$ and  we evaluate the $\sigma^*_a$ for each one. It might be that not all the four estimated values have the same compatibility with a Gaussian centered in zero since, due to the lack of configurations, one value might be overestimated. Increasing $\lambda$, the number of quadruplets for which the hypothesis test does not give the same answer for every estimated value decreases. We observed that the minimum $\lambda$ value for which there are no quadruplets that answer differently to the hypothesis test is very close, within less than an $8\%$ difference, to $\hat{\lambda}$.\\           
 \indent
\textit{PLM Decimation}
In the decimation procedure we maximize the total log-likelihood for all modes
\begin{eqnarray}
{\cal L}\equiv \frac{1}{N}\sum_{i}^N{\cal L}^{(0)}_i
\end{eqnarray}
starting from ${\cal L}_{\rm max}$ defined on a full graph and
inferring all the values of the $N_q = N(N-1)(N-2)(N-3)/24\sim N^4$ couplings on that
network. Sorting the couplings by their absolute value and taking away ({\em decimating})
the $N_0$ smallest we are left with a network of $N_q-N_0 = x N_q$ non-zero
couplings. A new inference iteration, including minimization of ${\cal
  L}(x)$ and sorting, will allow to decimate further another group of
the smallest couplings. And so on and so forth, as far as the
couplings of the decimated model network are more than the couplings
of the true system. Indeed, as far as the number of inferred
parameters is larger than the true number of parameters the pseudolikelihood
is not expected to change and will stick to its maximum value ${\cal
  L}_{\rm max}$ (overfitting). The indication of the true number of
couplings will be given, indeed, right by the $x$ value of the fraction of non-decimated couplings across which
${\cal L}$ starts decreasing because the number of inferred parameters
is less than the real number of parameters (underfitting).  The
minimum conceivable ${\cal L}(x)$ is for the non-interacting system:
${\cal L}_{\rm min}= {\cal L}(0)$. To enhance the determination of the
true network connectivity $x^*$, below which statistical
interpolation becomes less reliable, Decelle and Ricci-Tersenghi
\cite{Decelle14} introduced the {\em tilted} pseudolikelihood ${\cal
  L}_t$
\begin{eqnarray}
\nonumber
{\cal L}_t\equiv{\cal L}(x)-x {\cal L}_{\rm max}-(1-x){\cal L}_{\rm min}
\end{eqnarray}
such that 
\begin{eqnarray}
&&{\cal L}_t(0)={\cal L}_t(1)=0
\nonumber \\
\nonumber
&&{\cal L}_t(x\leq x^*) \simeq x({\cal L}_{\rm max}-{\cal L}_{\rm min}).
\end{eqnarray}
We stress that using the total likelihood, cf. Eq. (\ref{eq:Ltot}), the symmetry of coupling constants under indices permutation is automatically enforced.\\
\textbf{Random Graphs} We report the distributions of connectivities in the various graphs that we used to test the reconstruction methods exposed in the main manuscript for the complex spherical model (SM), cf.  Eq. (\ref{eq:H}), and the XY model, cf. Eq. (\ref{eq:HXY}), at finite $N$.
In the XY models the $x$ axis is the connectivity per variable.
In the  SM  the total number of couplings scales like $N^3$: 
the $x$ axis represents, in this case, the connectivity per variable node rescaled by $N^2$.

   \begin{table}[t!]
\begin{center}
\begin{tabular}{|c|c|c||c|c|}
\hline
$N$ & phasor ER & phasor ML & XY ER & XY ML\\
\hline
$16$ & $N_q=366$ & $N_q=252$ & $N_q=16$ & $N_q=47$ \\
\hline
$32$ & $N_q=2949$ & $N_q=2360$ & $N_q=32$ & $N_q=72$\\
\hline
$128$ & $N_q= 188742$ & $N_q= 168672$ & $N_q=128$ & $N_q=275$\\
\hline
\end{tabular}
\caption{Total number of quadruplets in all simulated networks used to yield data to be tested by our pseudolikelihood inference methods. }
\label{tab}
\end{center}
\end{table}
   
     \begin{figure}[t!]
   	\centering
   	\includegraphics[width=1\linewidth]{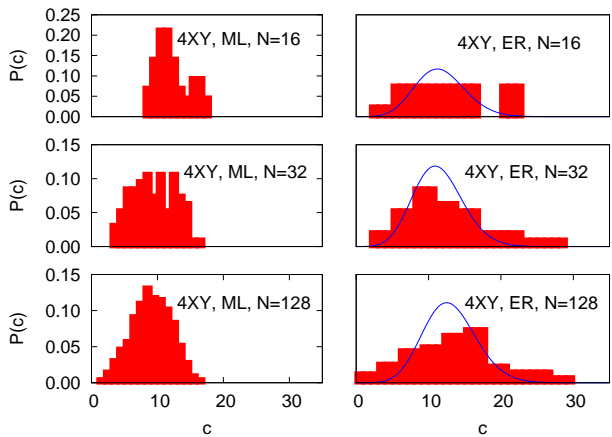}
   	\caption{	Distribution of connectivity per variable node of the $4$XY models on ER-like (left) and Mode-Locking-like (right) sparse graphs of  sizes $N=16,32,128$. The blue curves in the right panels are Poisson distributions with same mean as the empirical distributions.}
   	\label{fig:conn_XY}
   \end{figure} 
   
   \begin{figure}[t!]
   	\centering
   	\includegraphics[width=1\linewidth]{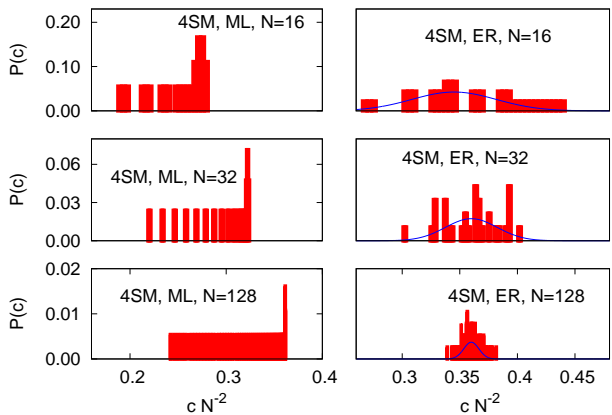}
   	\caption{ Distribution of rescaled connectivity $c/N^2$ in the $4$-phasor models - termed "spherical model" (SM)  in the plots -  on ER-like and ML-like dense ($O(N^3)$) graphs of sizes $N=16,32,128$. The blue curves in the right panels are Poisson distributions with same mean as the data derived distributions.}
   	\label{fig:conn_SM}
   \end{figure} 
   
\begin{figure}[t!]
   	\centering
   	\includegraphics[width=1\linewidth]{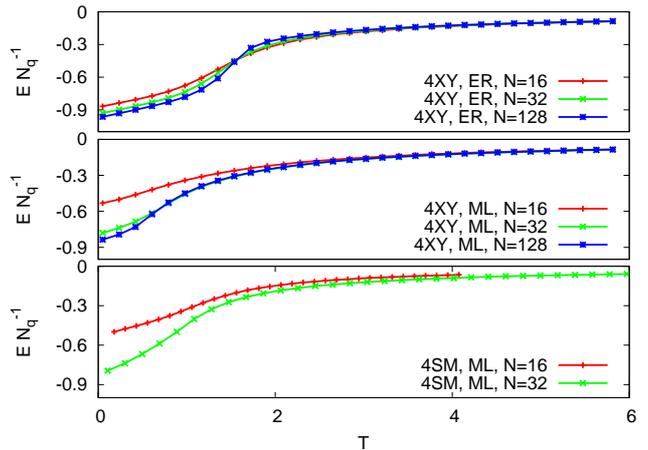}
   	\caption{Internal energy vs. temperature in various simulated instances of the models of Eqs. (\ref{eq:H},\ref{eq:HXY}) }
   	\label{fig:energy}
   \end{figure}

  The distributions in the Erdos-Renyi-like graphs and in the Mode-Locked-like graphs in pairwise interacing systems are known to  tend  to Poissonian distributions in the thermodynamic limiti $N\to\infty$ \cite{MarruzzoTesi}, though the ML-graph convergence is much slower for increasing $N$.
  As a comparison,  in the right panels of Figs. \ref{fig:conn_XY}, \ref{fig:conn_SM} we plot the relative Poissonian distributions with the same average. 
  Finite size effects are clearly strong for the small simulated sizes.
The actual number of quadruplet couplings $N_q$ for each simulated instance is reported for all considered models in Tab. \ref{tab}.\\
\textbf{Finite size critical temperature}
We refer several times to a critical region in $T$ for which inference
works better that at higher or lower $T$. In particular, the
reconstruction error is up to one order of magnitude smaller and the
network topology is correctly reconstructed with the need of less
configurations data. In Fig. \ref{fig:energy} we display the energy
plots of the models used to produce equilibrium data with Exchange
Monte Carlo simulations. In the thermodynamic limit $N \to \infty$
they should display a true critical point behavior where a phase
transition occurs from a paramagnetic-like phase to a phase-locked or
to a spin-glass phase. For the $4$XY model on ER sparse graphs these
temperatures are analytically known, as $T_c=0$ for $N_q/N=1$ and
$T_c>0$ for $N_q/N=2$ \cite{Marruzzo16}.  For finite sizes the
mathematical discontinuity is approximated by a steep, though smeared,
descent as $T$ decreases. This step becomes steeper and steeper as $N$
increases, eventually reaching the limit of a true singularity
representing a true phase transition.  Indeed, for finite systems, no
actual phase transition occurs. However, one can identify finite size
pseudocritical points that, for systems of increasing size $N$, form a
series converging to the true critical point. Operatively, we take as
finite size critical points $T_c(N)$ the peak of the specific heat $d
E/d T$ from $E_N(T)$ data in Fig. \ref{fig:energy}.  For the simulated
systems whose energy is displayed in Fig. \ref{fig:energy} the
$T_c(N)$ are reported in Tab. \ref{tab:tc}.

   \begin{table}[t!]
\begin{center}
\begin{tabular}{|c|c||c|c|}
\hline
$N$ & phasor ML & XY ER & XY ML\\
\hline
$16$ & $1.07(1)$ & $1.34(1)$ & $0.50(1)$ \\
\hline
$32$  & $0.91(1)$ & $1.39(1)$ & $0.72(1)$\\
\hline
$128$  &  & $1.53(1)$ & $0.63(1)$\\
\hline
\end{tabular}
\caption{Finite size proxies for the critical point. }
\label{tab:tc}
\end{center}
\end{table}


%

\noindent
\textbf{Authors Contributions}
A.M., P.T., F.A.,  A.P. and L.L. work on the programs solving the inverse problems and on the data analysis as well as on the Monte Carlo simulations and on the graphs analysis. A.M., P.T., F.A.,  A.P. and L.L. reviewed and approved the final version of this manuscript.\\     
\noindent
\textbf{Additional Information}\\
\noindent 
\textbf{Competing financial interests} The authors declare no competing financial interests.\\ 
\noindent
\textbf{Acknowledgements}\\
We thank  Federico Ricci Tersenghi for fruitful discussions. This project has received funding from the European Research Council (ERC) under the European Union's Horizon 2020 research and innovation program, Project {\em LoTGlasSy}, Grant Agreement No. 694925 and from the Italian Ministry for Education, University and Research (MIUR) under the PRIN 2015, Project {\em Statistical Mechanics and Complexity}, CINECA code 2015K7KK8L\_005.  

\end{document}